\documentclass[conference]{IEEEtran}
\IEEEoverridecommandlockouts

\usepackage{cite}
\usepackage{amsmath,amssymb,amsfonts}
\usepackage{algorithmic}
\usepackage{graphicx}
\usepackage{graphics}       
\usepackage{epsfig}
\usepackage{textcomp}
\usepackage{xcolor}
\def\BibTeX{{\rm B\kern-.05em{\sc i\kern-.025em b}\kern-.08em
    T\kern-.1667em\lower.7ex\hbox{E}\kern-.125emX}}

\usepackage{subcaption}
\usepackage{import}
\usepackage{svg}
\usepackage{mathtools} 

\begin{document}

\title{Robust two-degrees-of-freedom control of hydraulic drive with remote wireless operation}



\author{Riccardo Checchin, Michael Ruderman, Roberto Oboe  
\thanks{R. Checchin was with Department of Management and Engineering,
University of Padova, Vicenza, Italy}
\thanks{M. Ruderman is with Department of Engineering Sciences, University of Agder (UiA).
Postal address: P.B. 422, Kristiansand,4604, Norway. \newline
Correspondence to:
        {\tt\small michael.ruderman@uia.no}}
\thanks{R. Oboe is with Department of Management and Engineering,
University of Padova, Vicenza, Italy}
\thanks{This work was partially supported by the ERASMUS+ program.}
\thanks{\textcolor[rgb]{0.00,0.00,1.00}{Accepted manuscript, presented at IEEE ICM 2023 conference}}
}

\maketitle

\bstctlcite{references:BSTcontrol}

\begin{abstract}
In this paper, a controller design targeting the remotely operated
hydraulic drive system is presented. A two-degrees-of-freedom PID
position controller is used, which is designed so that to maximize
the integral action under robust constraint. A linearized model of
the system plant, affected by the parameters uncertainties such as
variable communication time-delay and overall system gain, is
formulated and serves for the control design and analysis. The
performed control synthesis and evaluation are targeting the
remote operation where the wireless communication channel cannot
secure a deterministic real-time of the control loop. The provided
analysis of uncertainties makes it possible to ensure system
stability under proper conditions. The theoretically expected
results are confirmed through laboratory experiments on the
standard industrial hydraulic components.
\end{abstract}


\section{Introduction}  

Remote wireless operation, with networked control systems, become
more crucial in various industries, especially due to more and
more spatially distributed sensors, actuators, and different-level
controllers, see e.g. a survey provided in
\cite{park2017wireless}. The associated communication constraints,
often expressed in varying transmission delays, see e.g.
\cite{heemels2010networked}, will remain one of the inevitable
stumbling block on the way of guaranteeing the stability and
desirable level of performance of the feedback control. The latter
has to be robust to specific variations in the delays and
transmission intervals. Already in early two-thousands, the
practical investigations were made in remote control of the
mechatronic systems over communication networks, see e.g.
\cite{delay13}. At the same time, one can notice the recent works
e.g. \cite{skogestad2018should} which prioritize a robust PID
design over more sensitive (to uncertainties) compensators of the
time-delays, like for example the Smith predictor. The hydraulic
drives, often as integrated or even embedded mechatronic systems,
are becoming frequently operated in the remote and distributed
plants, also via teleoperation and networked control, see e.g.
\cite{banthia2017lyapunov}. Having mostly a relatively high level
of the process and measurement noise and, at the same time, being
safety-critical for several (often outdoor and harsh-environment)
applications, also with large forces and corresponding payloads,
the hydraulic drives are particularly demanding for a stable
operation under the control delays and uncertainties.

In this paper, one aims to investigate a robust
two-degrees-of-freedom (2DOF) PID position controller, following
the methodology provided in \cite{astrom6} and adapting it to a
complex hydraulic drive system with substantial time-delays in the
control loop. Here it is worth emphasizing that our goal is not to
obtain the best implementation of the remote control, so that to
reduce much as possible the RTT (round trip time) in the feedback
loop. Instead, one is interested rather in setting up a remotely
controlled system with relatively large and variable RTT that
makes the system stability more challenging. Moreover, the
hydraulic systems are subject to considerable nonlinearities, as
it is well known from the subject literature, see e.g. in
\cite{Hanif14,hydraulic2020}. This fact makes a robust linear
control design particulary challenging and relevant for the
practical applications. Here, one purposefully uses the simple
2DOF PID control structure, see e.g. \cite{astrom3}, due to its
wide spread and acceptance in industries. The recent work provides
experimental tests in a laboratory setting, while real industrial
hydraulic components like servo-valve and linear cylinder are in
use. For more details on the experimental laboratory testbed, the
reader is referred to \cite{hydraulic2,hydraulic3}, while the
preliminary and detailed results, which are building fundament for
the present work, can be found in \cite{Checchin22}.

The rest of the paper is organized as follows. Section II provides
the control-oriented system modeling while emphasizing: (i) the
main steps of a system linearization, (ii) its most crucial
uncertainties (including time-delays), and (iii) the resulted
perturbed model serving the robust control design approach.
Section III is dedicated to the 2DOF PID control design. The
experimental results are reported in section IV, while the brief
conclusions are drawn in section V.

\section{Control oriented system modeling}

\subsection{Linearized model}

First, one analyzes the hydraulic system at hand. Starting from
the more physical constitutive equations, a control-oriented
linear system model is obtained. The hydraulic system is composed
of a directional control valve (DCV), i.e. servo-valve, and a
cylinder. The DCV has a nonlinear behavior due to the dead-zone
and saturation of the spool stroke. Also nonlinear friction force
appears when moving the piston inside the cylinder chambers. The
reduced-order nonlinear model, which results from the full-order
nonlinear model (see \cite{hydraulic1} for details), constitutes
the basis for system linearization.

\begin{figure}[htbp]
    \centering
    \includegraphics[width=0.9\columnwidth]{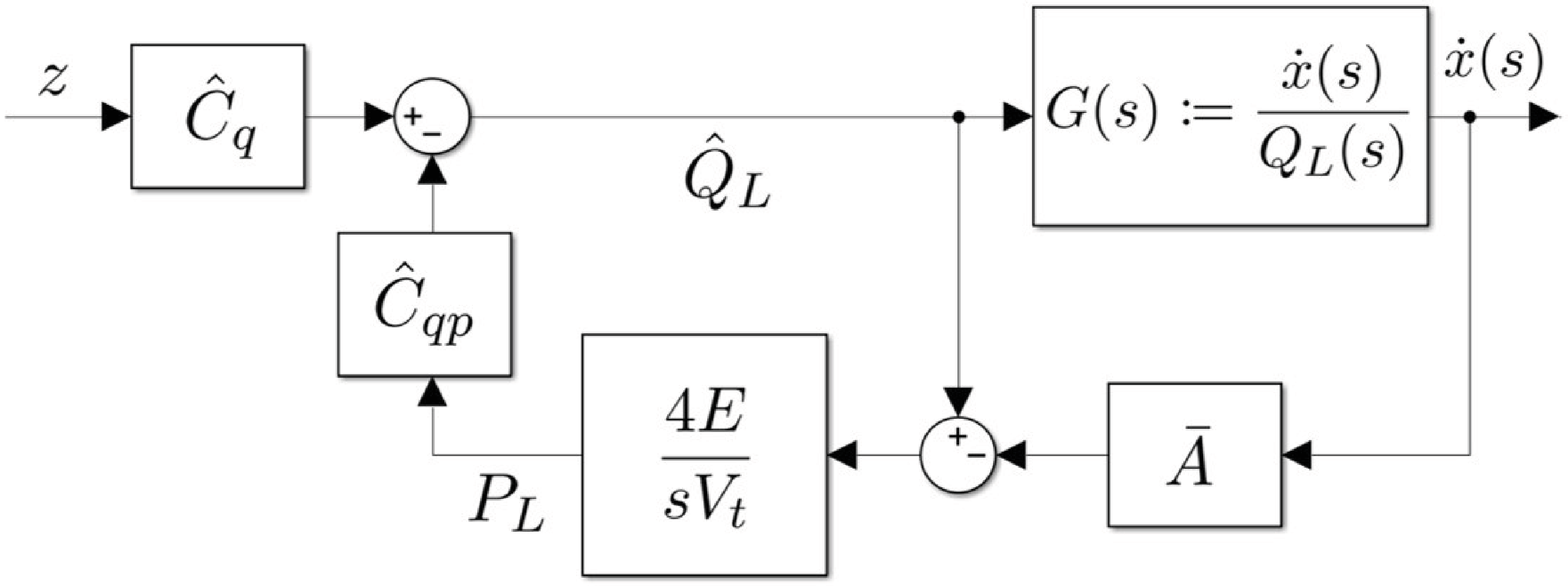}
    \caption{Block diagram of the linearized process model.}
\label{fig:blockLin}
\end{figure}

From the reduced-order model of~\cite{hydraulic1} one has the
equivalent orifice equation
\begin{equation}
    Q_L=zK\sqrt{\frac{1}{2}(P_S-\operatorname{sign(z)}P_L)}\text{ ,}
    \label{eq:Qleq}
\end{equation}
where $Q_L$ is the volumetric flow rate of hydraulic medium, $K$
is the flow coefficient (i.e. which is a valve's constant), $P_S$
is the supply pressure, $P_L$ is the system load pressure, and $z$
is an internal variable related to the controlled state of the
DCV. Recall that $z$ is subject to the dead-zone and saturation
nonlinearities. One linearizes \eqref{eq:Qleq} for the fixed
$\hat{P}_L, \hat{z}$ using the chain rule. Following to that, one
obtains
\begin{equation}
    \hat{Q}_L=\hat{C}_q z-\hat{C}_{qp}\hat{P}_L,
    \label{eq:Clinear}
\end{equation}
with
\begin{equation}
    \hat{C}_q \coloneqq \frac{\partial Q_L}{\partial z}\bigg|_{\hat{P}_L}=K\sqrt{\frac{1}{2}(P_S-\operatorname{sign(z)}\hat{P}_L)}\text{ ,}
    \label{eq:Cq}
\end{equation}
and
\begin{equation}
    \hat{C}_{qp} \coloneqq -\frac{\partial Q_L}{\partial P_L}\bigg|_{\hat{z}}=\frac{\hat{z}K\operatorname{sign(\hat{z})}}{4\sqrt{\frac{1}{2}(P_S-\operatorname{sign(\hat{z})}P_L)}}\text{ .}
    \label{eq:Cqp}
\end{equation}
In Fig.~\ref{fig:blockLin}, one can see the block diagram of the
linearized system model, from the internal control state $z$ to
the output speed $\dot{x}$ of the cylinder rod.
In the block diagram of Fig.~\ref{fig:blockLin}, the following
transfer function is defined
\begin{equation}
    G(s) \coloneqq \frac{\dot{x}(s)}{Q_L(s)}
    = \frac{1}{(ms+\sigma_{lin}) V_t (4\bar{A}E)^{-1}  s +\bar{A}}\text{,}
    \label{eq:G(s)}
\end{equation}
where $m$ is the total moving mass, $\sigma_{lin}$ is the
equivalent viscous friction coefficient, $V_t$ is the total oil
volume flowing in the hydraulic circuit, $E$ is the bulk modulus,
and $\bar{A}$ is the mean area of the working piston surface of
the hydraulic cylinder. Self-evident is that $s$ represents the
complex Laplace variable. From the block diagram of
Fig.~\ref{fig:blockLin}, one can directly compute the transfer
function from $Q_L$ to $P_L$ as
\begin{equation}
    R(s) \coloneqq \frac{P_L(s)}{Q_L(s)}
        = \frac{4E(1-G(s)\bar{A})}{sV_t}
        = (ms + \sigma_{lin})G(s)\bar{A}^{-1}\text{ .}
    \label{eq:R(s)}
\end{equation}
Hence, the transfer function from $z(s)$ to $\dot{x}(s)$ results
in
\begin{equation}
    \hat{G}(s) \coloneqq \frac{\dot{x}(s)}{z(s)} = \frac{\hat{C}_qG(s)}{1+\hat{C}_{qp}R(s)}      \text{ .}
    \label{eq:Ghat(s)}
\end{equation}

\begin{figure}[!h]
    \centering
    \includegraphics[width=0.7\columnwidth]{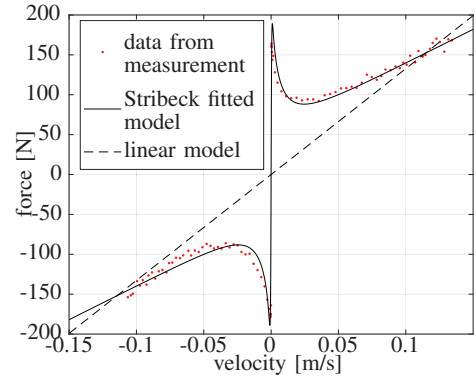}
    \caption{Comparison between experimental and model data.}
    \label{fig:friction}
\end{figure}

Important to notice is that $\sigma_{lin} >0$ is used in
\eqref{eq:G(s)} and \eqref{eq:R(s)} in order to describe the
linear viscous friction behavior. An appropriate choice of
$\sigma_{lin}$ appears relevant because the real frictional
behavior is strongly nonlinear, as confirmed with identification
data shown below. Indeed, one can recognize in the velocity-force
coordinates in Fig.~\ref{fig:friction}, the experimentally
collected data points \cite{hydraulic2} for both motion
directions. Note that the points are determined each from the
steady-state measurement of relative velocity $\dot{x}$ and load
pressure at the unidirectional drive experiments with a nearly
constant speed. Recall that at steady-state, the load pressure is
proportional to the total resistive force, which is due to the
total friction when no other forces are in balance. The
experimental data are also fitted with the Stribeck friction
model, see e.g. \cite[eq. (2)]{ruderman2019virtual}, which is then
used for the sake of a more accurate numerical simulation. The
dashed line in Fig.~\ref{fig:friction} reflects the selected
linear friction coefficient $\sigma_{lin}$. This is chosen so that
the linear model reaches the same friction force level towards the
end of the considered velocity range. At the same time, for the
maximal possible relative velocity
$\dot{x}_{max}=0.25[\frac{m}{s}]$, detected in the open-loop
experiments, the $\sigma_{lin}$ constitutes a trade-off which
minimizes an integral square error between the identified Stribeck
and linear viscous friction model.

Now, the obtained linear model~\eqref{eq:Ghat(s)} is augmented by
the overall communication time delay $\tau$. Therefore, one
derives the overall process transfer function, i.e. from the
delayed control channel $z$ to the output position of interest
$x$, as
\begin{equation}
    P(s) \coloneqq \frac{x(s)}{z(s)} e^{-s\tau} = \hat{G}(s) s^{-1} e^{-s\tau}\text{ .}
        \label{eq:P(s)}
\end{equation}
Since the overall plant model has two gaining factors related to
the linearization, cf. \eqref{eq:Cq}, \eqref{eq:Cqp}, and one
weakly-known time delay factor, one has to deal with a
\emph{nominal process transfer function}, for which one fixes the
nominal parameter values $\tau_{\text{nom}}$, $C_{q,\text{nom}}$,
$C_{qp,\text{nom}}$. Later, those parameters are considered as a
source of uncertainties in the robust control design. The nominal
process transfer function is given explicitly by
\begin{equation}
\begin{split}
    P_{\text{nom}}(s)
        = \frac{ e^{-s\tau_{\text{nom}}}\frac{C_{q,\text{nom}}\bar{A}}{\sigma_{lin}C_{qp,\text{nom}}+\bar{A}^2} } { s\biggl[s^2\frac{mV_t}{4E(\sigma_{lin}C_{qp,\text{nom}}+\bar{A}^2)}+s\frac{C_{qp,\text{nom}}m+\frac{\sigma_{lin}V_t}{4E}}{\sigma_{lin}C_{qp,\text{nom}}+\bar{A}^2}+1\biggr] }\text{ .}
\end{split}
        \label{eq:P_nom(s)}
\end{equation}

\subsection{Uncertainties analysis}

As shown above, the linearized gaining factors $C_q$ and $C_{qp}$
depend on the input and state operating points $\hat{z}$ and
$\hat{P}_L$, respectively. On the other hand, the communication
delay can be of a purely stochastic nature, and the only upper
bound $\tau_{\max}$ can be assumed. In order to assign the nominal
values of the uncertain parameters, consider the following.

\begin{itemize}

\item In Fig.~\ref{fig:Cqp}, the $C_{qp}(P_L,z)$ surface is shown
for the sake of visualization. Since the $C_{qp}$-surface is not
uniformly distributed but symmetrical with respect to the
$(P_L,z)$-origin, one proper choice of the nominal value is the
integral mean value over the whole operative ranges, i.e. $z \in
(-1, +1)$ and $P_L \in (-P_s, +P_s)$. As it is visible from eq.
\eqref{eq:Cqp}, $\forall z \in [-1,1] \lim_{P_L \to P_s}
C_{qp}(P_L,z) \to +\infty$, so that one can stop integration at
e.g. 95 \% of $P_s$, which is an appropriate range for the load
pressure $P_L$. In a similar way as $C_{qp}$, one can then assert
the value for $C_q$.
\begin{figure}[!h]
    \centering
    \includegraphics[width=0.85\columnwidth]{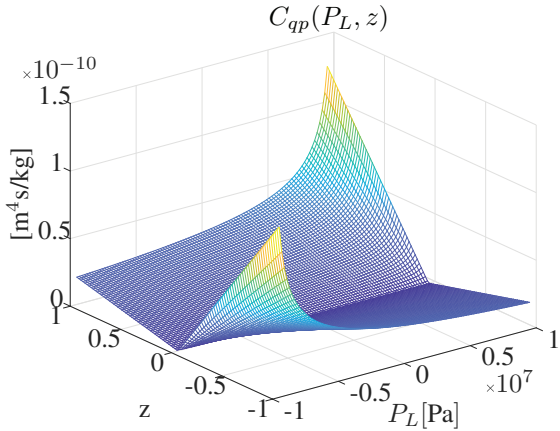}
    \caption{Variations of the gaining factor $C_{qp}(P_L,z)$.}
    \label{fig:Cqp}
\end{figure}

\item In order to analyze the (RTT) delay $\tau$, more than
10 different communication experiments on the hardware of
experimental testbed were performed. The spatial distance between
the routers of a wireless network and the intensity of the
exchanged data were varied. In Fig.~\ref{fig:gamma10control}, a
typical distribution determined from the measurements of $\tau$ is
shown. Note that the 0.01 [s] column-width of the histogram
corresponds the 0.01 [s] quantization of $\tau$, the latter due to
the communication time period captured on side of the remote
controller. As can be found in the literature, see e.g.
\cite{delay0}, \cite{delay13}, an RTT model for TCP/IP
communication often suggests the $\gamma$-distribution. From
Fig.~\ref{fig:gamma10control}, one can recognize that the fitted
$\gamma$-distribution is well in accord with the measurement.
Based on the determined $\gamma$-distribution, the nominal value
$\tau_{\text{nom}}$ is calculated as average by evaluating the
integral mean value.
\begin{figure}[!h]
   \centering
    \includegraphics[width=0.7\columnwidth]{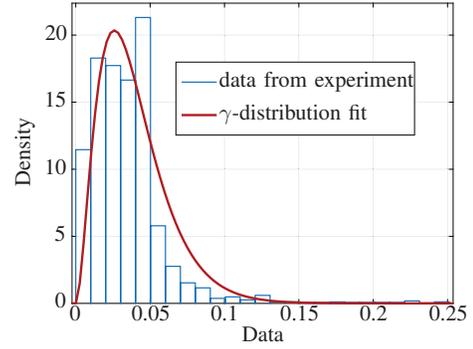}
    \caption{\emph{$\gamma$-distribution} fit of the measured $\tau$ distribution.}
    \label{fig:gamma10control}
\end{figure}

\end{itemize}

\subsection{Perturbed system model}

The perturbed system model is defined as following
\begin{equation}
P_{\tau}(i\omega)=\big[1+\Delta_U(i\omega)W_U(i\omega)\big]P_{\text{nom}}(i\omega)\text{ ,}
\label{eq:mulUncer}
\end{equation}
where $W_U(i\omega)$ is a fixed stable transfer function for
weighing the uncertainties, and $\Delta_U(i\omega)$ is a variable
stable transfer function satisfying
$\|\Delta_U(i\omega)\|_{\infty}<1$. The complex number and angular
frequency are denoted by $i$ and $\omega$, correspondingly. The
used structure model is very well known as \emph{multiplicative
perturbation}. Indeed, the transfer characteristics
$P_{\tau}(i\omega)$ could additionally vary from the nominal one
in a range defined by $\Delta_U(i\omega)W_U(i\omega)$ in the
magnitude, and by $\Delta_U(i\omega)$ in the phase. One chooses
the multiplicative perturbation structure to represent the
uncertain system because of the following reasoning. From eq.
\eqref{eq:P_nom(s)}, one can recognize that the nominal transfer
function has a pair of conjugate-complex poles and a free
integrator. The transfer function parameters, related to the pole
pair, are the gaining factor, natural frequency, and damping
coefficient
\begin{equation}
    k(\hat{C}_q, \hat{C}_{qp}) = \frac{\hat{C}_q\bar{A}}{\sigma_{lin}\hat{C}_{qp}+\bar{A}^2}\text{ ,}
    \label{eq:K_P(s)}
\end{equation}
\begin{equation}
    \omega_n(\hat{C}_{qp}) = \sqrt{\frac{4E(\sigma_{lin}\hat{C}_{qp}+\bar{A}^2)}{mV_t}}\text{ ,}
    \label{eq:w_P(s)}
\end{equation}
\begin{equation}
    \xi(\hat{C}_{qp})=\frac{\hat{C}_{qp}m+\frac{\sigma_{lin}V_t}{4E}}{\sigma_{lin}\hat{C}_{qp}+\bar{A}^2} \cdot \frac{\omega_n}{2}\text{ .}
    \label{eq:csi_P(s)}
\end{equation}
respectively. Following to that, one can compute the maximum
relative deviation for these three parameters, which are varying
subject to linearization. This is done by iterative calculation of
\eqref{eq:K_P(s)}, \eqref{eq:w_P(s)}, \eqref{eq:csi_P(s)} over the
range of possible $(C_{qp},C_q)$ values, followed by a comparison
with the parameter values obtained for the nominal
$(C_{qp,\text{nom}},C_{q,\text{nom}})$. This results in
\begin{equation}
\begin{split}
\text{max relative deviation }\Bigl[k(C_{qp},C_q)\Bigr]&= 71.37\%\text{ ,}\\
\text{max relative deviation }\Bigl[\omega_n(C_{qp},C_q)\Bigr] & = 6.1\%\text{ ,}\\
\text{max relative deviation }\Bigl[\xi(C_{qp},C_q)\Bigr] & = 25.1\%\text{ .}\\
\end{split}
    \label{eq:relErrK}
\end{equation}
One can recognize that the maximal possible deviation on
$\bigl[k(C_{qp},C_q)\bigr]$ is significantly larger comparing to
the maximal deviations of $\bigl[\omega_n(\hat{C}_{qp})\bigr]$ and
$\bigl[\xi(\hat{C}_{qp})\bigr]$. Such dominance in the gain
variation justifies the multiplicative perturbation assumption
made for the present system.

A sufficient robust stability condition for the model with the
multiplicative perturbation, following to e.g. \cite{doyle1}, is
\begin{equation}
\|W_U(i\omega)T(i\omega)\|_{\infty}<1\text{ ,}
\label{eq:RSC}
\end{equation}
where $T(i\omega)$ is the closed-loop transfer function. One can
easily see that the robust stability \eqref{eq:RSC} depends on the
weighting function $W_U(i\omega)$. Since it has to satisfy the
condition \eqref{eq:mulUncer}, one can write
\begin{equation}
    \Bigl | \frac{P_{\tau}(i\omega)}{P_{\text{nom}}(i\omega)} - 1 \Bigr | \leq \bigl | W_U(i\omega) \bigr |\text{ .}
    \label{eq:Wu}
\end{equation}
Since the multiplicative perturbation differs the $P_{\tau}$ and
$P_{\text{nom}}$ transfer functions by exactly the gaining factor
and time-delay element, the inequality \eqref{eq:Wu} can be
transformed into, cf. \cite{doyle1},
\begin{equation}
    \bigl | k e^{ - \tau i \omega} - 1 \bigr | \leq \bigl | W_U(i\omega) \bigr |\text{ .}
    \label{eq:Wu1}
\end{equation}
In order to satisfy \eqref{eq:Wu1} for all possible $C_q$,
$C_{qp}$, and $\tau$ in the range of variations, consider the
available upper bounds $k_{\max}$ and $\tau_{max}$. Then, one
needs to find a stable transfer function $W_U(i\omega)$ which can
guarantee the inequality \eqref{eq:Wu1} holds for the
$(k_{\max},\tau_{max})$ pair, so that it remains valid also for
all $\quad 0 < k < k_{\max}$ and $0 < \tau < \tau_{\max}$.
Assuming the second-order lead transfer function
\begin{equation}
    W_U(s)=k_w\frac{\frac{s^2}{\omega_z^2}+2\xi_z\frac{s}{\omega_z}+1}{\frac{s^2}{\omega_p^2}+2\xi_p\frac{s}{\omega_p}+1}\text{ ,}
    \label{eq:Wu(s)}
\end{equation}
as the weighting function for robust stability \eqref{eq:RSC}, the
parameter values $\omega_{z,p}$, $\xi_{z,p}$, and $k_w$ are fitted
by minimizing the square error between $W_U(i\omega)$ and the
left-hand-side of \eqref{eq:Wu1} for a certain range of angular
frequencies around both corner frequencies $\omega_{z} <
\omega_{p}$ of the lead element \eqref{eq:Wu(s)}.

\section{Control design}    

The control design follows the methodology provided in
\cite{astrom3}, \cite{astrom6}. The control objectives to be
achieved are:
\begin{enumerate}
\item fast load disturbance response;
\item robustness against the model uncertainties;
\item robustness against the measurement noise;
\item set point response accuracy.
\end{enumerate}
The control solution includes the following measures:
\begin{enumerate}
\item maximize the integral action of the PID controller;
\item define a robust constraint;
\item apply the second-order filter $F_n(s)$ to feedback;
\item tune the set point parameter $b$ and filter $F_{\text{sp}}(s)$.
\end{enumerate}

\subsection{Robust feedback design}

The applied design algorithm \cite{astrom3} maximizes the integral
action $k_i$ of the standard PID feedback controller
\begin{equation}
    C_c(s) = k_p + \frac{k_i}{s} + s k_d\text{ ,}
    \label{eq:C}
\end{equation}
satisfying the constraint
\begin{equation}
    f(k_p,k_i,k_d,\omega) = \Big|\big[k_p+i(k_d\omega-k_i/\omega) \big]P_{\text{nom}}(i\omega)+1\Big|^2 \geq r^2\text{ .}
    \label{eq:f>r2}
\end{equation}
The inequality \eqref{eq:f>r2} guarantees that the open-loop
transfer function $C_c(i\omega)P_{\text{nom}}(i\omega)$ lays
outside the circle of the radius $r$, which is centered
in$(-1,i0)$ of the complex plane. Then, it is the Nyquist theorem
which guarantees the stability of the closed-loop system under
this condition. Therefore, the system remains stable until
$C(i\omega)P_{\text{nom}}(i\omega)$ surrounds $(-1,0)$. The
parameter $r$ describes the robustness of the system and
\eqref{eq:f>r2} is called a robust constraint.

If looking at the definition of the infinite norm of the
sensitivity function $S$, cf. \cite{sigurd1}, which is the maximum
absolute value of $S(i\omega)$ over all frequencies $\omega$, i.e.
\begin{equation}
    M_s \coloneqq \max_{\omega}|S(i\omega)| = \max_{\omega} \biggl| \frac{1}{1+P_{\text{nom}}C_c(i\omega)} \biggr|\text{ ,}
    \label{eq:Ms}
\end{equation}
one can write $r=\frac{1}{M_s}$. Then, one can choose a value of
$M_s$ according to the robustness required from the feedback
control system. One assumes $M_s=1.1$, cf. \cite{astrom3}, in
accord with a relatively high level of perturbations in the
system.

\begin{figure*}[htbp]
    \centering
    \includegraphics[width=0.65\textwidth]{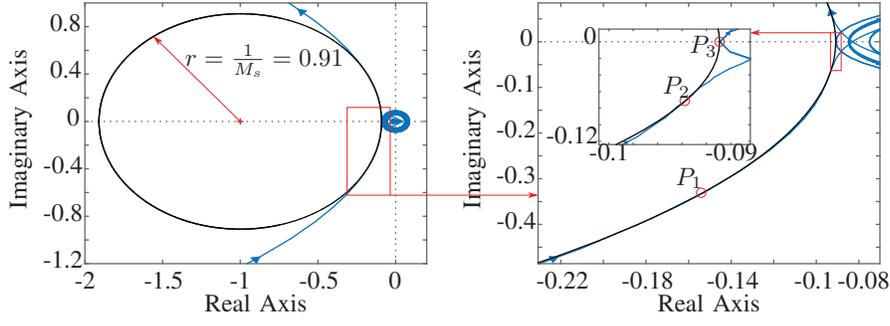}
    \caption{Nyquist diagram of the open loop transfer function with PID controller designed with \eqref{eq:kPID}.}
    \label{fig:nyquistSol}
\end{figure*}

For the identified nominal process
\begin{equation}
    P_{\text{nom}}(s)
        = e^{-0.03s}\frac{8.255\cdot 10^5}{s\bigl[ s^2+948s+2.219\cdot 10^6 \bigr]}\text{ ,}
        \label{eq:Pnom(s)_exp}
\end{equation}
one obtains the following control gains
\begin{equation}
\begin{split}
k_{p}&=12.7534\text{ ,}\\
k_{i}&=31.1783\text{ ,}\\
k_{d}&=0.1472\text{ .}
\end{split}
\label{eq:kPID}
\end{equation}
by applying the $k_i$ maximization algorithm proposed in
\cite{astrom3}. In Fig. \ref{fig:nyquistSol}, one can verify the
Nyquist plot of the open-loop transfer function with respect to
the above robust constraint. In particular, one can see that the
algorithm maximizes the PID parameters until reaching the robust
constraint, i.e. by touching the $r$-circle at three points $P_1$,
$P_2$, $P_3$, i.e. at three different angular frequencies. The
resulted robust feedback control, designed for the nominal
process, can be further assessed on the robust stability
\eqref{eq:RSC}. Here, the closed-loop transfer function
$T(i\omega) =
P_{\text{nom}}C_c(i\omega)\bigl(1+P_{\text{nom}}C_c(i\omega)\bigr)^{-1}$
is multiplied with $W_U(i\omega)$, which is satisfying
\eqref{eq:Wu1} for $\tau = \tau_{\max}$. Worth noting is that
while $\tau_{\text{nom}}=0.03$, its upper bound
$\tau_{\max}=0.11$, cf. Fig. \ref{fig:gamma10control}, is still
satisfying the robust stability \eqref{eq:RSC}.

\subsection{2DOF control structure}

\begin{figure}[htbp]
\centering
    \includegraphics[width=1\columnwidth]{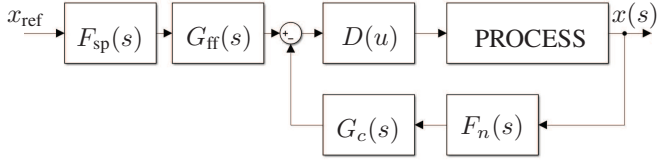}
    \caption{Block diagram of the 2DOF control structure.}
    \label{fig:fullControl}
\end{figure}

In Fig. \ref{fig:fullControl}, the overall 2DOF control structure
is shown with
\begin{equation}
    G_{\text{ff}}(s) = bk_p+\frac{k_i}{s}\text{ ,}
    \label{eq:Gff}
\end{equation}
\begin{equation}
    F_{\text{n}}(s) = \frac{1}{(1+\frac{s}{2N\omega_o})^{2}}\text{ ,}
    \label{eq:Fn}
\end{equation}
which are the feed-forward and feed-back filters, respectively.
Here $b$ is the set-point parameter to be tuned, and $N$ is the
parameter of the noise cutoff frequency to be adjusted for having
a suitable feedback response. $N = 5$ is set as an appropriate
value, cf. \cite{astrom6}. $D(u)$ is the inverse map of the
orifice dead-zone, cf. \cite{hydraulic2}, which has to compensate
the static non-linearity which was not considered in the nominal
process $P_{\text{nom}}(s)$.
Further, compute the pre-filter, cf. \cite{astrom6},
\begin{equation}
    F_{\text{sp}}(s)=\frac{1}{1+s\frac{2\pi}{\omega_{\text{sp}}}\sqrt{M_w^2-1}}=\frac{1}{1+s\tau_{\text{sp}}}\text{ ,}
    \label{eq:Fsp(s)}
\end{equation}
so as to guarantee that $\max_\omega \bigl[|W(i\omega)|\bigr] \leq
1$, where $M_w=\max_\omega[|W(i\omega)|]$, and
\begin{equation}
    W(s)=\frac{x(s)}{x_{\text{ref}}(s)}=
    \frac{G_{\text{ff}}P_{\text{nom}}(s)}{1+G_{\text{c}}F_{\text{n}}P_{\text{nom}}(s)}\text{ .}
    \label{eq:W(s)}
\end{equation}
This way, the overall closed-loop transfer function, from the
reference position to the output rod position, will have $M_w=1$,
hence, ensuring no overshoot in the step response.

\section{Experimental results}   

The robust 2DOF controller, developed according to sections II and
III, is experimentally evaluated on the laboratory testbed
\cite{hydraulic2}.
\begin{figure}[!h]
\centering
\includegraphics[width=0.54\columnwidth]{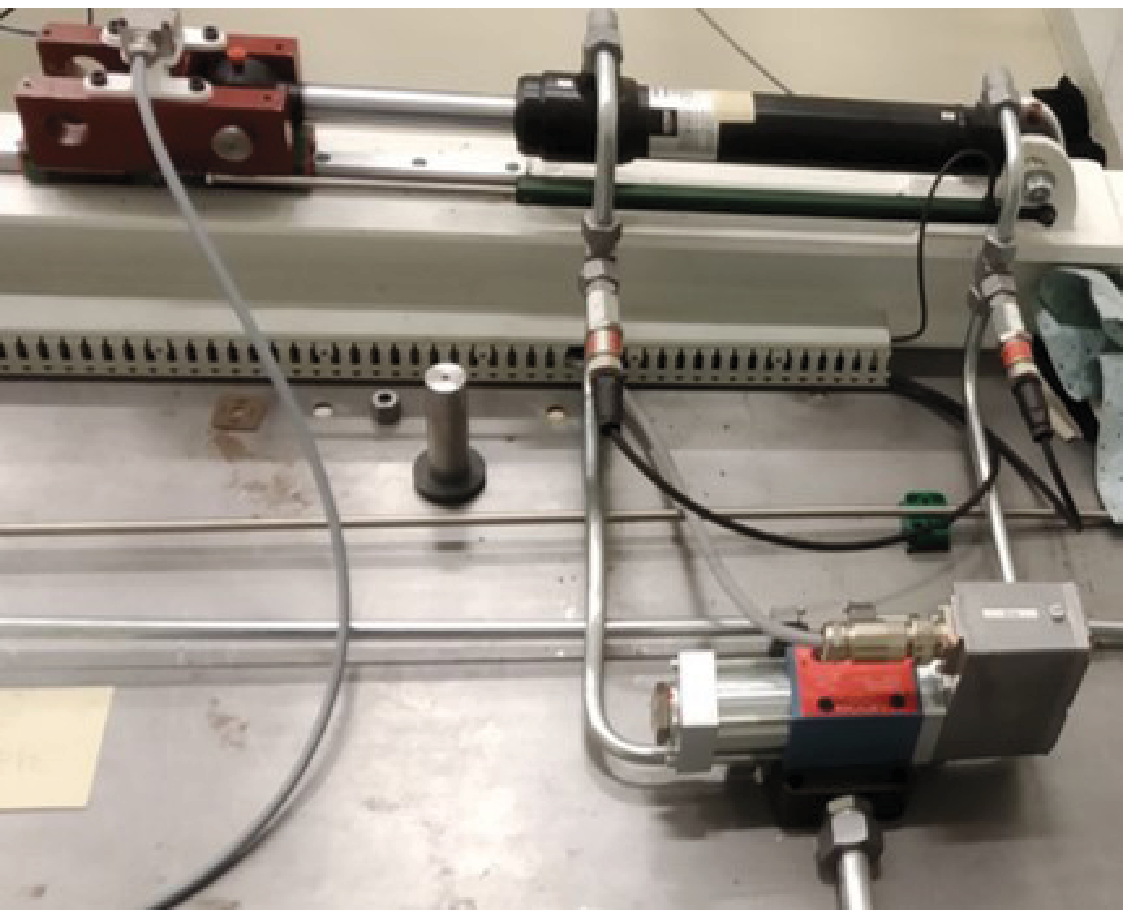}
\hspace{0.2cm}
\includegraphics[width=0.33\columnwidth]{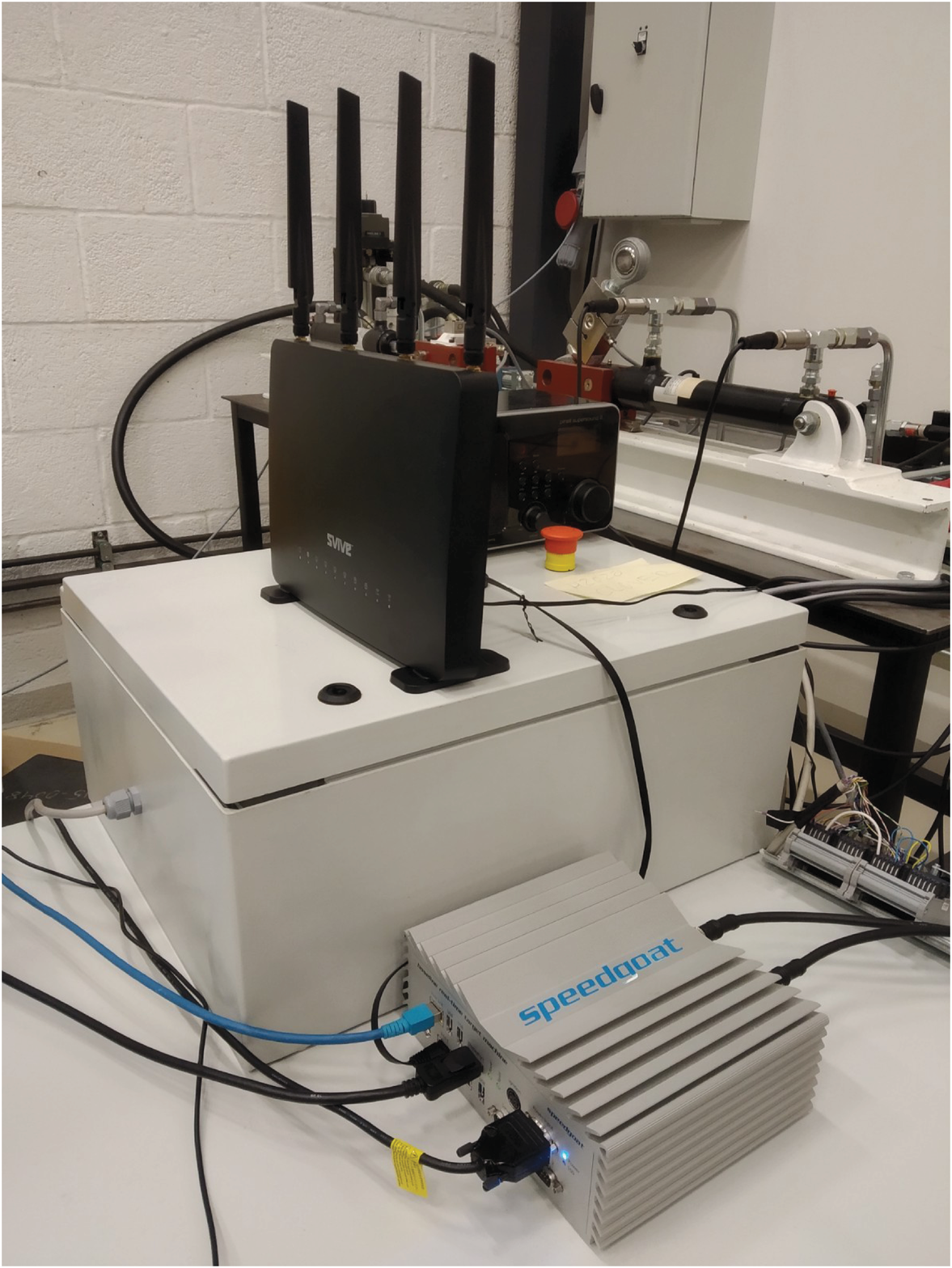}
\caption{Laboratory experimental setup: hydraulic testbed
\cite{hydraulic2} on the left; switch cabinet with the embedded
hardware interfaces, real-time board, and WiFi routers on the
right.} \label{fig:setup}
\end{figure}
The used experimental setup is shown in Fig. \ref{fig:setup}. The
real-time SpeedGoat board is controlling the power and sensor
interfaces, with deterministic sampling time set to 0.0005 sec.
The SpeedGoat board is connected to a standard WiFi router as a
point-to-point TCP/IP communication, this way sending the measured
output values, received from the sensor, and receiving the control
values from a remote PC-based controller. The remote PC-based
controller is realized on a standard conventional laptop computer.
Due to non-real-time processes of the TCP/IP-based socket and
implemented control on the running PC, the minimal possible time
delay $\tau_{\min}=0.01$ sec appears as a communication sample
time.

Two communication scenarios have been evaluated: (i) the remote
control communication performs via a point-to-point Ethernet
connection between the real-time SpeedGoat and PC-based
controller; (ii) the remote control communication performs via a
wireless connection between the real-time SpeedGoat and PC-based
controller by means of WiFi. Note that for (ii), various spatial
distances between the WiFi communicating nodes were tested. In
both communication scenarios, the RTT delay, which corresponds to
the overall $\tau$, was monitored and recorded in the feedback
loop.

\begin{figure}[!h]
\centering
    \begin{subfigure}[c]{1\columnwidth}
        \includegraphics[width=0.98\columnwidth]{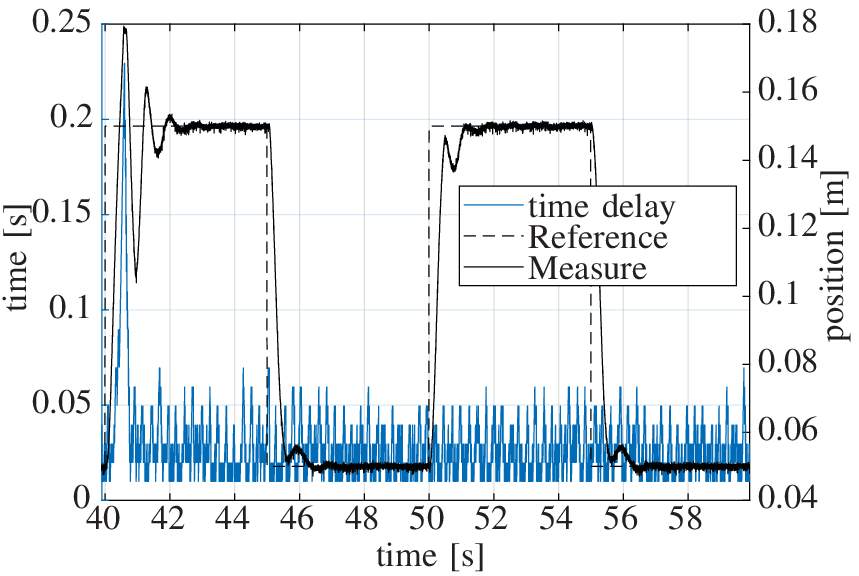}
    \end{subfigure}\vspace{1mm}
    \begin{subfigure}[c]{1\columnwidth}
        \includegraphics[width=0.98\columnwidth]{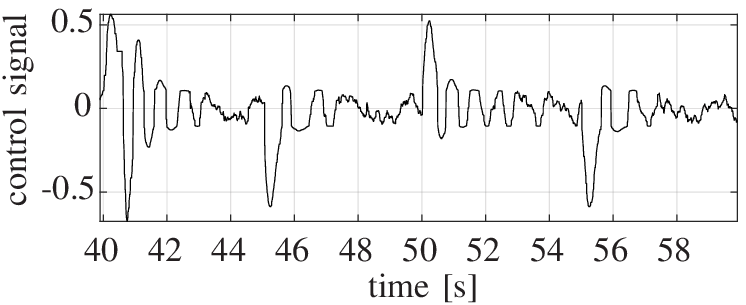}
    \end{subfigure}
    \caption{Experimental control response (Ethernet connection).}
    \label{fig:test1}
\end{figure}
\begin{figure}[!h]
\centering
    \includegraphics[width=0.98\columnwidth]{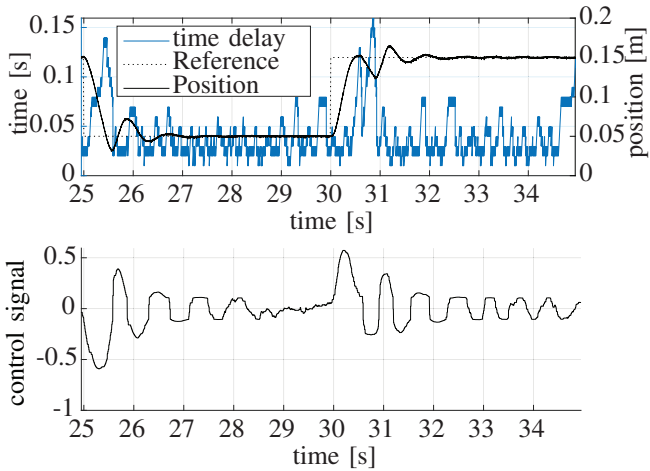}
    \caption{Experimental control response (WiFi connection).}
    \label{fig:test4sim8}
\end{figure}
\begin{figure}[!h]
\centering
    \includegraphics[width=0.98\columnwidth]{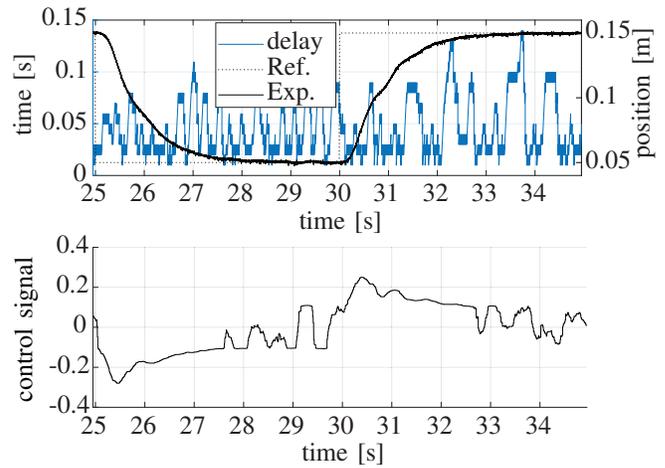}
    \caption{Experimental control response, using additionally the set-point filter \eqref{eq:Fsp(s)} (WiFi connection).}
    \label{fig:test5sim7}
\end{figure}

The results shown below visualize the reference and output
position of the hydraulic cylinder, together with the recorded
time delay in the upper plots, and the corresponding control
signal in the lower plots. A series of square-pulse-shaped (with
0.1 m amplitude) positioning experiments are shown for
communication scenario (i) in Fig. \ref{fig:test1}. Note that for
time-delay values higher than 0.11 sec, cf. sections II and III,
an instability can occur. Further it is noted that the smoothing
set-point filter \eqref{eq:Fsp(s)} is not applied at that stage.
The same type positioning experiments for the communication
scenario (ii) are shown in Fig. \ref{fig:test4sim8}. Note that
despite $\tau$ is transiently exceeding the $\tau_{\max}$ value,
the robustly designed 2DOF PID control maintains a stable and
performant response, with solely shortcoming of transient
oscillations.
In Fig. \ref{fig:test5sim7}, the communication scenario (ii) is
shown when additionally applying the set-point filter
\eqref{eq:Fsp(s)}, cf. Fig. \ref{fig:fullControl}. This way, the
square-pulse-shaped reference $x_{\text{ref}}$ is additionally
lagged, which slows down the overall output response but, in
return, avoids the transient oscillations. Worth noting is that
the increased spatial distance between the WiFi nodes led to the
larger $\tau$-variations in this case, with more frequent values
closer to $\tau_{\max}$, cf. Figs. \ref{fig:test1},
\ref{fig:test4sim8}, and \ref{fig:test5sim7}.

\section{Conclusions}   

This paper shows the application of the linear 2DOF control
methodology \cite{astrom3} to the nonlinear uncertain hydraulic
drive system with significant time delays due to the wireless
remote control. The robust constraint on the varying system gain
and time delay parameters were specified and used for the feedback
control design that meets the robust stability criteria for
multiplicative disturbances. It is demonstrated in the laboratory
experiments, accomplished on the standard industrial components,
that the robust 2DOF PID control is well applicable to remotely
operated hydraulic drives with only output measurement of the
cylinder stroke.


\bibliographystyle{IEEEtran}
\bibliography{bibliography}

\begin{thebibliography}{10}
\providecommand{\url}[1]{#1}
\csname url@rmstyle\endcsname
\providecommand{\newblock}{\relax}
\providecommand{\bibinfo}[2]{#2}
\providecommand\BIBentrySTDinterwordspacing{\spaceskip=0pt\relax}
\providecommand\BIBentryALTinterwordstretchfactor{4}
\providecommand\BIBentryALTinterwordspacing{\spaceskip=\fontdimen2\font plus
\BIBentryALTinterwordstretchfactor\fontdimen3\font minus
  \fontdimen4\font\relax}
\providecommand\BIBforeignlanguage[2]{{%
\expandafter\ifx\csname l@#1\endcsname\relax
\typeout{** WARNING: IEEEtran.bst: No hyphenation pattern has been}%
\typeout{** loaded for the language `#1'. Using the pattern for}%
\typeout{** the default language instead.}%
\else
\language=\csname l@#1\endcsname
\fi
#2}}

\bibitem{park2017wireless}
P.~Park, S.~C. Ergen, C.~Fischione, C.~Lu, and K.~H. Johansson, ``Wireless
  network design for control systems: A survey,'' \emph{IEEE Communications
  Surveys \& Tutorials}, vol.~20, no.~2, pp. 978--1013, 2017.

\bibitem{heemels2010networked}
W.~M.~H. Heemels, A.~R. Teel, N.~Van~de Wouw, and D.~Ne{\v{s}}i{\'c},
  ``Networked control systems with communication constraints: Tradeoffs between
  transmission intervals, delays and performance,'' \emph{IEEE Transactions on
  Automatic control}, vol.~55, no.~8, pp. 1781--1796, 2010.

\bibitem{delay13}
O.~Roesch and H.~Roth, ``Remote control of mechatronic systems over
  communication networks,'' in \emph{IEEE International Conference Mechatronics
  and Automation}, 2005, pp. 1648--1653.

\bibitem{skogestad2018should}
C.~Grimholt and S.~Skogestad, ``Should we forget the smith predictor?''
  \emph{IFAC-PapersOnLine}, vol.~51, no.~4, pp. 769--774, 2018.

\bibitem{banthia2017lyapunov}
V.~Banthia, K.~Zareinia, S.~Balakrishnan, and N.~Sepehri, ``A {Lyapunov} stable
  controller for bilateral haptic teleoperation of single-rod hydraulic
  actuators,'' \emph{Journal of Dynamic Systems, Measurement, and Control},
  vol. 139, no.~11, p. 111001, 2017.

\bibitem{astrom6}
H.~Panagopoulos, K.~Astrom, and T.~Hagglund, ``Design of {PID} controllers
  based on constrained optimisation,'' \emph{IEE Proceedings -- Control Theory
  and Applications}, vol. 149, pp. 32--40, 2002.

\bibitem{Hanif14}
H.~Chaudhry, \emph{Applied Hydraulic Transients}, 3rd~ed.\hskip 1em plus 0.5em
  minus 0.4em\relax Springer, 2014.

\bibitem{hydraulic2020}
N.~Manring and R.~Fales, \emph{Hydraulic Control Systems}, 2nd~ed.\hskip 1em
  plus 0.5em minus 0.4em\relax John Wiley and Sons, 2020.

\bibitem{astrom3}
K.~J. Åstr\"om and T.~H\"agglund, \emph{Advanced PID Control}.\hskip 1em plus
  0.5em minus 0.4em\relax ISA - the instrumentation, systems, and automation
  society, 2006.

\bibitem{hydraulic2}
P.~Pasolli and M.~Ruderman, ``Linearized piecewise affine in control and states
  hydraulic system: Modeling and identification,'' in \emph{IEEE 44th Annual
  Conf. of Indust. Elect. Society (IECON)}, 2018, pp. 4537--4544.

\bibitem{hydraulic3}
P.~Pasolli and M.~Ruderman, ``Hybrid position/force control for hydraulic
  actuators,'' in \emph{IEEE 28th Mediterranean Conference on Control and
  Automation (MED)}, 2020, pp. 73--78.

\bibitem{Checchin22}
\BIBentryALTinterwordspacing
R.~Checchin, ``Robust design of a remote motion control system for hydraulic
  mechatronic drive with wireless operation,'' 2022, {Master} thesis. [Online].
  Available: \url{https://hdl.handle.net/11250/2985189}
\BIBentrySTDinterwordspacing

\bibitem{hydraulic1}
M.~Ruderman, ``Full- and reduced-order model of hydraulic cylinder for motion
  control,'' in \emph{IEEE 43rd Annual Conference of the IEEE Industrial
  Electronics Society (IECON)}, 2017, pp. 7275--7280.

\bibitem{ruderman2019virtual}
M.~Ruderman, L.~Fridman, and P.~Pasolli, ``Virtual sensing of load forces in
  hydraulic actuators using second-and higher-order sliding modes,''
  \emph{Control Engineering Practice}, vol.~92, p. 104151, 2019.

\bibitem{delay0}
A.~Vespignani and R.~Pastor-Satorras, \emph{Evolution and Structure of the
  Internet}.\hskip 1em plus 0.5em minus 0.4em\relax Cambridge University Press,
  2004.

\bibitem{doyle1}
J.~Doyle, B.~Francis, and A.~Tannenbaum, \emph{Feedback Control Theory}.\hskip
  1em plus 0.5em minus 0.4em\relax Dover Publications Inc, 1990.

\bibitem{sigurd1}
S.~Skogestad and I.~Postlethwaite, \emph{Multivariable feedback control:
  analysis and design}, 2nd~ed.\hskip 1em plus 0.5em minus 0.4em\relax John
  Wiley \& sons, 2005.

\end{thebibliography}

\end{document}